\documentstyle[12pt,epsf,rotate]{article}
\input FEYNMAN
\newcommand{\pr}{\paragraph{}}
\newcommand{\nn}{\nonumber}
\newcommand{\be}{\begin{equation}}
\newcommand{\ee}{\end{equation}}
\newcommand{\bea}{\begin{eqnarray}}
\newcommand{\eea}{\end{eqnarray}}

\begin{document}
\newcommand{\nd}[1]{/\hspace{-0.5em} #1}
\begin{titlepage}

\begin{flushright}
NTUA-65/97 \\
OUTP-97-36P \\

hep-lat/9707027 \\
\end{flushright}

\begin{centering}
\vspace{.05in}
{\Large {\bf Dynamical Gauge Symmetry Breaking 
and Superconductivity in three-dimensional
systems
\\}}
 
\vspace{.1in}
 K. Farakos 
 
\vspace{.05in}

National Technical University of Athens, 
Physics Department,
Zografou Campus
GR-157 73, Athens, Greece, \\

\vspace{.05in}
and \\
\vspace{.05in}
N.E. Mavromatos$^{*}$ \\
\vspace{.05in}
University of Oxford, Department of (Theoretical) Physics, 
1 Keble Road OX1 3NP, Oxford, U.K. \\

\vspace{.2in}
{\bf Abstract} \\
\vspace{.05in}
\end{centering}
{\small We discuss 
dynamical breaking
of non-abelian 
gauge groups in three dimensional (lattice) gauge systems
via the formation of fermion condensates. 
A physically relevant example,
motivated by condensed-matter physics, is that of 
a fermionic gauge theory with  
group 
$SU(2)\otimes U_S(1) \otimes U_{E}(1)$.
In the strong $U_S(1)$ region, the $SU(2)$ symmetry breaks 
down to a $U(1)$, due to the formation 
of a parity-invariant fermion condensate.  
We conjecture a phase diagram for the theory involving 
a critical line, which separates the regions
of broken $SU(2)$ 
symmetry
from those where the symmetry is restored. 
In the broken phase, the effective Abelian gauge 
theory  is
closely related to an earlier model 
of
two-dimensional parity-invariant 
superconductivity in doped antiferromagnets.
The superconductivity in the model occurs  
in the Kosterlitz-Thouless mode, since 
strong phase fluctuations prevent the existence of a 
local order parameter.
Some physical consequences 
of the $SU(2) \times U_S(1)$ phase diagram 
for the (doping-dependent) parameter space of this 
condensed-matter model are briefly discussed.}

\vspace{.4in}

\begin{flushleft} 
July 1997 \\
$^{(*)}$~P.P.A.R.C. Advanced Fellow. \\
\end{flushleft} 

\end{titlepage}

\section{Introduction}

Gauge symmetry breaking without an elementary Higgs particle,
which proceeds via the dynamical formation of fermion condensates,  
has been a fascinating idea, which however had been tried
rather inconclusively, so far, in attempts to understand 
either chiral 
symmetry breaking in four-dimensional QCD via a new strong 
interaction (technicolour)~\cite{xsb}, or in 
the breaking of a local gauge symmetry through the formation 
of pair condensates in non-singlet channels~\cite{dimop}.
In all such scenaria the basic idea is that there exists an 
energy scale at which the gauge coupling becomes strong enough 
so as to favour the formation of non-zero fermion condensates
$<{\bar f} f'>$ which are not invariant under the 
global or local symmetry in question. 
It is the purpose of this short note to point out that 
similar scenaria of dynamical gauge symmetry breaking 
in three-dimensional gauge theories~\cite{farak} lead to interesting 
and unconventional superconducting properties of the theory
after coupling to electromagnetism~\cite{dor},  
and therefore may be 
of interest to condensed matter community.
\pr
In a recent publication~\cite{fm} we 
have argued that the doped 
large-$U$ Hubbard (antiferromagnetic) models possess
a {\it hidden} local {\it non-Abelian} $SU(2)\times U_S(1)$ 
phase symmetry related to spin interactions.
This symmetry was discovered through a generalised 
{\it slave-fermion} ansatz for {\it spin-charge 
separation}~\cite{Anderson}, 
which allows intersublattice 
hopping for holons, and hence spin flip~\footnote{Non-abelian 
gauge symmetry structures 
for doped antiferromagnets,
in a formally different context though, i.e. by employing slave-boson
techniques,
have also been proposed by other authors~\cite{leewen}. 
However, the patterns of symmetry breaking 
discussed here, and in ref. \cite{fm},  
are physically different from those 
approaches, and they allow for a unified description 
of slave-boson and slave-fermion approaches 
to spin-charge separation.}. 
The spin-charge separation 
may be physically interpreted as implying an effective 
`substructure' of the electrons due to the 
many body interactions in the medium. 
This sort of idea, originating from Anderson's 
RVB theory of spinons and holons~\cite{Anderson},
was also pursued recently by 
Laughlin, although from a (formally at least)
different perspective~\cite{quark}. 
\pr
The effective long wavelength 
model of such a statistical system 
is remarkably similar to a 
three-dimensional gauge model 
of particle physics proposed in ref. \cite{farak}
as a toy example for chiral symmetry breaking 
in QCD. In that work, it has been argued that {\it dynamical 
generation} of a fermion mass gap 
due to the 
$U_S(1)$ subgroup of $ SU(2) \times U_S(1)$
breaks the $SU(2)$ subgroup down to a $\sigma_3-U(1)$ 
group, where $\sigma_3$ is the $2\times 2$  Pauli matrix. 
{}From the particle-theory view point 
this is a Higgs mechanism without an {\it elementary} 
Higgs excitation. 
The analysis carries over to the condensed-matter case,
if one associates the mass gap to the holon condensate~\cite{fm}. 
The resulting effective theory of the light degrees of freedom 
is then similar to the 
continuum limit of \cite{dor}
describing unconventional parity-conserving superconductivity.  
\pr
Apart from their condensed-matter applications, however, 
we believe that such patterns of symmetry breaking 
are also of interest to the particle-physics community.
For instance, it is known that high-temperature 
gauge theories in four dimensions become effectively 
three-dimensional 
euclidean systems. Therefore, one cannot exclude the possibility 
that the scenaria discussed here, and in refs. \cite{farak,dor}, 
might be of relevance to this case in the future.  
For this reason we consider it as useful 
to expose the particle-physics community to 
the above ideas through this short note. 
\pr
The structure of the article is as follows: in sec. 2 we review briefly
the dynamical symmetry breaking patterns of $SU(2) \times U_S(1)$ model 
on the lattice. In sec. 3 we discuss the phase diagram of the theory
and argue in favour of the existence of a critical line separating 
the broken-$SU(2)$ phase from that where the symmetry is restored. 
In sec. 4 we present the superconducting properties of the 
broken phase upon coupling the system to electromagnetic potentials. 
Finally in section 5, instead of conclusions, we discuss very briefly
the application of these ideas to a specific model 
in condensed matter physics, which might 
have some relevance to the physics of high-temperature 
superconductors. The interested reader may find details on the physics 
of this model in ref. \cite{fm}.

\section{Dynamical Non-Abelian Gauge Symmetry  
Breaking on the Lattice}

To understand the nature of the non-Abelian gauge symmetries
under consideration it is instructive to start 
with the simpler case of a local Abelian gauge theory,
invariant under a group $U_S(1)$, which  
has a global $U(2)$ symmetry. Eventually we shall 
gauge this non-Abelian symmetry and make contact with the 
lattice models~\cite{fm}. 
We, therefore, begin by considering
the following three-dimensional 
continuum lagrangian~\cite{farak,fm}: 
\be 
{\cal L} = -\frac{1}{4}(F_{\mu\nu})^2 
+ {\overline \Psi} D_\mu\gamma _\mu \Psi -m_0 {\overline \Psi} \Psi 
\label{contmodel}
\ee
where $D_\mu = \partial _\mu -ig_1 a_\mu^S $,
and $F_{\mu\nu}$, is the corresponding field
strength
for the abelian (termed `statistical') gauge field $a^S_\mu$.
In the statistical model of ref. \cite{fm}, the field $a_S$ is responsible
for fractional statistics of the pertinent (holon) excitationson 
of the planar doped antiferromagnet. 
Anticipating the 
connection of (\ref{contmodel}) with the (naive) continuum 
limit of an appropriate stastistical (lattice) system, 
we may take the  
fermions $\Psi$ to be four-component spinors, 
due to lattice doubling. It is in this formalism 
that the global $U(2)$ symmetry will be constructed~\cite{farak,fm}. 
The
parity-conserving bare-mass $m_0$ term 
has been added by hand 
to facilitate Monte-Carlo studies~\cite{koutsoumbas} of 
dynamically-generated fermion masses 
as a result of the formation of fermion condensates
$<{\overline \Psi} \Psi >$ by the strong 
$U_S(1)$ coupling. The $m_0=0$ limit should be taken 
at the end.
The $\gamma _\mu $, $\mu =0,1,2$,  
matrices 
span the reducible $4 \times 4$ representation of the 
Dirac algebra in three dimensions in a fermionic theory with an 
{\it even} number of fermion flavours~\cite{app}:
$\gamma ^0=\left(\begin{array}{cc} {\bf \sigma}_3 
\qquad {\bf 0} \\{\bf 0} \qquad -{\bf \sigma}_3 \end{array} \right)
\qquad \gamma ^1 = \left(\begin{array}{cc} i{\bf \sigma}_1 
\qquad {\bf 0} \\{\bf 0} \qquad -i{\bf \sigma}_1 \end{array} \right), 
\gamma ^2 = \left(\begin{array}{cc} i{\bf \sigma}_2 
\qquad {\bf 0} \\{\bf 0} \qquad -i{\bf \sigma}_2 \end{array} \right)$
where ${\bf \sigma}$ are $2 \times 2$ Pauli matrices
and the (continuum) space-time is taken to have Minkowskian signature.
As well known~\cite{app} there exist two $4 \times 4 $ matrices 
which anticommute with $\gamma _\mu$,$\mu=0,1,2$,
and generate a `chiral' symmetry 
in a theory with even number of fermion species: 
$\gamma _3~=\left(\begin{array}{cc} 0 \qquad {\bf 1} \\
{\bf 1} \qquad 0 \end{array}\right), \qquad 
\gamma _5 =i\left(\begin{array}{cc} 0 \qquad {\bf 1} \\
{\bf -1} \qquad 0 \end{array}\right)$,
where the substructures are $2 \times 2$ matrices.
\pr
The set of generators 
\be
{\cal G} = \{ {\bf 1}, \gamma _3, \gamma _5, 
\Delta \equiv i\gamma_3\gamma _5 \}
\label{generators}
\ee
form~\cite{farak} 
a global $U(2) \simeq SU(2) \times U_E(1)$ symmetry. 
The identity matrix ${\bf 1}$ generates the $U_E(1)$ subgroup, 
while 
the other three form the SU(2) part of the group. 
In the statistical model for (magnetic) superconductivity 
of ref. \cite{fm}, which we shall describe 
briefly below, the global $U_E(1)$ corresponds to the electromagnetic
charge of the holons, and can be gauged by coupling the system to 
an external electromagnetic potential $A_\mu$.
\pr
In two-component notation for the spinors 
$\Psi _i$, the bilinears
\bea
&~&{\cal A}_1 \equiv -i[{\overline \Psi}_1 \Psi_2 
- {\overline \Psi}_2 \Psi_1] ,  
\qquad {\cal A}_2 \equiv {\overline \Psi}_1 \Psi_2 
+ {\overline \Psi}_2 \Psi_1 ,  
\qquad {\cal A}_3 \equiv {\overline \Psi}_1\Psi _1 
- {\overline \Psi}_2 \Psi_2 ,   
\nn \\
&~&{\cal F}_{1\mu} \equiv {\overline \Psi}_1\sigma _\mu \Psi _2 +
{\overline \Psi}_2\sigma_\mu \Psi_1, \qquad   
{\cal F}_{2\mu} \equiv i[{\overline \Psi}_1\sigma _\mu \Psi _2 -
{\overline \Psi}_2\sigma_\mu \Psi_1], \nn \\
&~&{\cal F}_{3\mu} \equiv {\overline \Psi}_1\sigma _\mu \Psi_1
-{\overline \Psi}_2\sigma _\mu \Psi_2,  
\label{triplets}
\eea
transform as {\it triplets} under $SU(2)$. 
\pr
On the other hand, 
the $SU(2)$ singlets are given by the bilinears: 
\be
{\cal A}_4 \equiv {\overline \Psi}_1\Psi_1
+ {\overline \Psi}_2\Psi_2, \qquad 
{\cal F}_{4,\mu} \equiv {\overline \Psi}_1\sigma _\mu \Psi_1
+ {\overline \Psi}_2\sigma _\mu \Psi_2~, \qquad \mu=0,1,2      
\label{singlets}
\ee
i.e. the singlets are the parity violating mass term, 
and the four-component fermion number. 
\pr
One may {\it gauge} the above group $SU(2)\times U_S(1) \times U_E(1)$, 
where, we remind the reader once again that  the electromagnetic symmetry $U_E(1)$ is gauged by 
coupling the system to external electromagnetic potentials: 
\be 
{\cal L}_2 \equiv =-\frac{1}{4}(F_{\mu\nu})^2 
-\frac{1}{4}({\cal G}_{\mu\nu})^2  +{\overline \Psi}D_\mu\gamma_\mu\Psi
-m_0{\overline \Psi}\sigma_3 \Psi 
\label{su2action}
\ee
where now $D_\mu = \partial_\mu -ig_1a_\mu^S-ig_2\sigma^aB_{a,\mu}
-\frac{e}{c}A_\mu$,
$B_\mu^a$ is the gauge potential of the local (`spin') $SU(2)$ group, 
and ${\cal G}_{\mu\nu}$ is the corresponding field strength. 
The fermions $\Psi$ are now, and in what follows, viewed as 
{\it two-component} spinors. 
Once we gauged the $SU(2)$ group, the colour structure is up and above 
the space-time Dirac structure, and in two-component 
notation the $SU(2)$ group is generated by the 
familiar $2 \times 2$ Pauli matrices $\sigma^a$, $a=1,2,3$. 
In this way, the fermion condensate ${\cal A}_3$
can be generated dynamically by means of a strongly-coupled
$U_S(1)$. In this context, energetics 
prohibits the generation of a parity-violating 
gauge invariant $SU(2)$ term~\cite{Vafa}, and so 
a parity-conserving mass term necessarily breaks~\cite{farak}
the $SU(2)$ group down to a $\sigma_3-U(1)$ sector~\cite{dor}, generated
by the $\sigma_3$ Pauli matrix in two-component notation. 
\pr
The above symmetry breaking patterns may be proven analytically~\cite{farak} 
on the lattice, in the strong $U_S(1)$ limit, $\beta_1 \rightarrow 0$.
The 
lattice lagrangian, corresponding to the continuum lagrangian 
(\ref{su2action}),
assumes the form: 
\bea
&~& S=\frac{1}{2}K \sum_{i,\mu}[{\overline \Psi}_i (-\gamma_\mu) 
U_{i,\mu}V_{i,\mu} \Psi_{i+\mu}  + \nonumber \\
&~& {\overline \Psi}_{i+\mu}
(\gamma _\mu)U^\dagger_{i,\mu}V^\dagger_{i,\mu}
\Psi _i ]  \nn \\
&~& + \beta _1 
\sum _{p} (1 - trU_p) + \beta _2 \sum _{p} (1-trV_p) 
\label{effeaction}
\eea
where $\mu =0,1,2$, $U_{i,\mu}=exp(i\theta _{i,\mu})$ 
represents the 
statistical $U_S(1)$ gauge field, 
$V_{i,\mu}=exp(i\sigma ^a B_a)$ is the $SU(2)$ gauge field, 
The fermions $\Psi $
are taken   
to be two-component (Wilson) spinors, in both Dirac and 
colour spaces~\cite{farak,fm}.
Here we have passed onto a three-dimensional 
Euclidean 
lattice formalism, in which ${\overline \Psi}$ is identified 
with $\Psi ^\dagger$.
In this convention the bilinears 
(\ref{triplets}),(\ref{singlets}) are hermitean quantities. 
\pr
In the strong-coupling limit, 
$\beta _1 =0$   
the field $U_S(1)$ field
may be integrated out analytically in the path integral
with the result
~\cite{farak}:
\be
Z = \int dV d{\overline \Psi}d \Psi exp(-S_{eff})
\label{action3}
\ee
where
\bea
&~&S_{eff} = \beta _2 \sum _{p} (1 -trV_p)
- \sum _{i,\mu}{\rm ln}I^{tr}_0(2\sqrt{y_{i\mu}}) \nn \\
&~& y_{i\mu} = -\frac{K^2}{4}tr[M^{(i)}(-\gamma _\mu)V_{i\mu}
M^{(i+\mu)}(\gamma_\mu)V_{i\mu}^\dagger] 
\label{action4}
\eea
where 
and 
\be
  M^{(i)}_{ab,\alpha\beta} 
\equiv \Psi _{i,b,\beta}{\overline \Psi}_{i,a,\alpha},~~a,b={\rm
colour},~\alpha,\beta={\rm Dirac},~i={\rm lattice~site}
\label{mesons}
\ee
are the meson states, and 
${\rm ln}I^{tr}_0$ denotes the logarithm of 
the zeroth order Modified Bessel function~\cite{Abra}, 
{\it truncated }
to an order determined by the number of the Grassmann (fermionic) 
degrees of freedom in the problem~\cite{kawamoto}. 
In our case, due to the $SU(2)$ and spin quantum numbers of the {\it 
lattice spinors} $\Psi$, one should retain terms in $-{\rm ln}I^{tr}_0$  
up to ${\cal O}(y^4)$:
\be
-{\rm ln}I^{tr}_0(2\sqrt{y_{i\mu}})=
-y_{i\mu}+\frac{1}{4}y^2_{i\mu}
-\frac{1}{9}y_{i\mu}^3 +\frac{11}{192}y_{i\mu}^4 
\label{trunclog}
\ee
The above expression 
is an {\it exact} result, irrespectively of the magnitude 
of $y_{i\mu}$. 
\pr
The low-energy (long-wavelength)
effective action is written as a path-integral 
in terms of gauge and meson fields,
$
Z=\int [dV dM]exp(-S_{eff}+ \sum_{i}tr{\rm ln}M^{(i)})$, 
where the meson-dependent term
comes from the Jacobian in passing from 
fermion integrals to meson ones~\cite{kawamoto}.
\pr
To identify the symmetry-breaking patterns of the 
gauge theory (\ref{trunclog})
one may concentrate on 
the lowest-order terms in $y_{i\mu}$, 
which will yield the gauge boson 
masses. Higher-order terms will describe interactions, as we shall discuss
in the next section. 
Keeping thus only the linear term in the expansion
yields~\cite{farak}:
\be
   {\rm ln}I^{trunc}_0(2\sqrt{y_{i\mu}})|_{linear} = 
y_{i\mu} =
-\frac{1}{4}K^2 tr[M^{(i)}(-\gamma _\mu)V_{i\mu}
M^{(i+\mu)}(\gamma _\mu)V_{i\mu}^\dagger]
\label{fkresult}
\ee
It is evident that symmetry-breaking patterns 
for $SU(2)$ will emerge out of a non zero VEV 
for the meson matrices $M^{(i)}$.
\pr
Lattice simulations of the model 
(\ref{effeaction}), with 
only a global $SU(2)$ symmetry, in the strong $U_S(1)$ coupling
limit $\beta _1=0$, and 
in the quenched approximation for fermions, 
have shown~\cite{koutsoumbas} that 
the states generated by the bilinears 
${\cal A}_1$ and ${\cal A}_2$ (\ref{triplets})
are {\it massless}, and therefore correspond to 
Goldstone Bosons, while the state generated 
by the bilinear ${\cal A}_3$ is massive.
The fact that 
members of the triplet SU(2) representation acquire
different masses is already evidence for 
symmetry breaking. 
To demonstrate this explicitly 
one uses  
the following expansion for the 
meson states in terms of the $SU(2)$
bilinears  (\ref{triplets}),(\ref{singlets})~\cite{farak}:
\bea
 &~& M^{(i)} = {\cal A}_3(i)\sigma_3 
+ {\cal A}_1(i)\sigma_1 + {\cal A}_2 (i)\sigma_2 + 
{\cal A}_4 (i){\bf 1} + \nn \\
&~& {\cal F}_{4,\mu}(i)\gamma ^\mu + {\cal F}_{1,\mu}(i)\gamma^\mu\sigma_1
+ {\cal F}_{2,\mu}(i)\gamma^\mu\sigma_2 + 
{\cal F}_{3,\mu}(i)\gamma ^\mu \sigma_3 
\label{mesontripl}
\eea
with $\mu=0,1,2$,
$\gamma_\mu$ are hermitean Dirac (space-time) $2\times 2$ matrices,
and $\sigma_a$, $a=1,2,3$ are the (hermitean) $2\times 2$ 
SU(2)-`colour' Pauli matrices. Note that 
the VEV of the matrix $<M^{(i)}>=u\sigma_3$
is proportional to the chiral condensate. 
Upon substituting (\ref{mesontripl})
in (\ref{fkresult}), taking into account that the 
SU(2) link variables may be expressed as 
$V_{i\mu} = cos(|{\bf B}_{i\mu}|) + i{\bf \sigma}.{\bf B}_{i\mu}
sin(|{\bf B}_{i\mu}|)/|{\bf B}_{i\mu}|$, 
and performing a naive perturbative expansion
over the fields ${\bf B}$ one finds:
\be
{\rm ln}I^{trunc}_0(2\sqrt{y_{i\mu}})|_{linear} \propto 
K^2 u^2
[(B^1_{i\mu})^2 + (B^2_{i\mu})^2]~+~{\rm interaction~terms}  
\label{massiveboson}
\ee
{}From this it follows that two of the $SU(2)$ gauge bosons, 
namely the $B^1$,$B^2$ become {\it massive}, with masses
proportional to the chiral condensate $u$: 
\be
{\rm B^{1,2}~boson~masses} \propto K^2u^2
\label{massbooson}
\ee
whilst the gauge boson $B^3$ remains {\it massless}.
Thus,
$SU(2)$ is broken down to a $U(1)$ subgroup, 
generated by the $\sigma_3$ Pauli matrix.

\begin{centering}
\begin{figure}[htb]
\epsfxsize=2in
\centerline{\epsffile{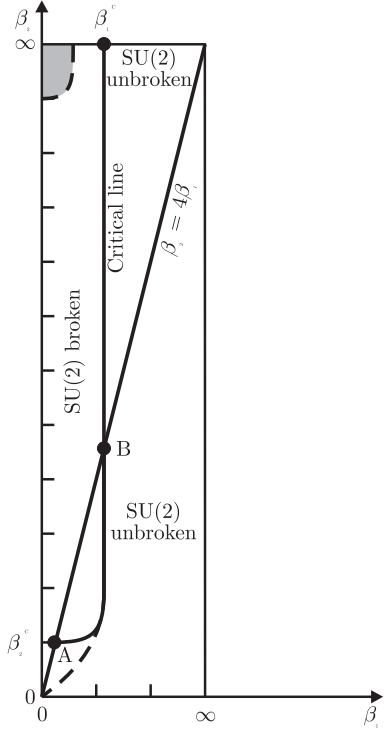}}
\vspace{1cm}
\caption{{\it Phase diagram 
for the $SU(2) \times U_S(1)$ gauge theory.
The critical line separates the phase
of (dynamically) broken $SU(2)$ symmetry 
from the unbroken phase. Its shape is conjectural 
at this stage, in particular with respect to 
the order of  magnitude
of $\beta_2^c$. The shaded region has been 
analysed in ref. \cite{farak}. 
The dashed line 
represents a probable critical line  
in the case of the statistical model of ref. \cite{fm}.
The straight line indicates the specific relation 
of the coupling constants in that model.}} 
\label{fig1}
\end{figure}
\end{centering}

\section{Phase Diagram of $SU(2) \times  U_S(1)$ Theory}

The reader might have noticed 
a similarity between the gauge fermion interaction terms
(\ref{fkresult}) 
and the corresponding Higgs-fermion interactions in the 
adjoint Higgs model~\cite{adjoint}. 
The adjoint Higgs model is characterized by a critical value of $K$ 
above which spontaneous symmetry breaking occurs.
In our case, however, as we shall argue below, the system is 
{\it always} in the {\it broken phase} in the region  
$\beta_2=\infty, \beta_1=0$, independently of $K$.  
\pr
To prove this, we consider the situation along the 
vertical axis of the phase diagram of 
fig. 1 at $\beta_1=0$. 
The effective potential 
depends on the variable $y$ defined in 
(\ref{action4}). To determine its form, 
in terms of the 
condensate, one uses the expansion (\ref{mesontripl}) 
and concentrates on the terms of the $SU(2)$ triplet, 
${\cal A}_i \equiv \Phi_i, i=1,2,3$. This triplet plays a r\^ole 
analogous to that of the Higgs triplet in the adjoint 
Higgs model~\cite{adjoint}. In that model, 
one of the members of the triplet, 
along the $\sigma_3$ direction of $SU(2)$, 
acquires a vacuum expectation value,
exactly as it happens in our case above. 
The form of the effective potential in the naive continuum limit
(lattice spacing $a \rightarrow 0$ ) and in the weak $SU(2)$ case
may be found from (\ref{trunclog}) 
by expanding the $SU(2)$ link variables $U_{ij} \simeq 1 + O(agA)$,
where $g$ is the group coupling, $a$ is the lattice spacing and 
$A$ is the gauge potential. 
Taking the limit $g\rightarrow 0$ 
($\beta_2 =\infty$) 
one then obtains $y(x) \sim 
\frac{K^2}{2}Tr\Phi^2(x)=K^2\sum_{i=1}^3 \Phi_i^2$.
The tree-level effective potential 
is then: 
\bea 
&~&V_{eff} = -3K^2R^2 + \frac{3}{4}K^4R^4 - \frac{1}{3}K^6R^6 
+ \frac{33}{192}K^8R^8 \nn \\
&~&R^2 \equiv \frac{1}{2}Tr\Phi^2=\sum_{i=1}^3 \Phi_i^2  
\label{smally}
\eea
The coefficient $3$ originates from the summation over the 
three link variables at each site of our three-dimensional 
lattice gauge theory. 
The potential (\ref{smally}) 
has the characteristic double-well shape of 
a symmetry-breaking (adjoint) Higgs potential, with 
a non-trivial minimum at $KR \simeq 1.3$ (c.f. figure 2).
Hence, for any value of $K$ there is always a corresponding 
Parity-invariant condensate, which is inversely 
proportional to $K$, implying that  
there is {\it no 
critical value of } $K$ above which the symmetry changes.
The system always remains in 
a single (broken $SU(2)$ symmetry) 
phase in the region $\beta_1=0, \beta_2 ={\rm large}$, irrespectively
of the strength of $K$. 
The reason for this behaviour, in contrast to the adjoint Higgs
case~\cite{adjoint}, where a critical $K$ {\it does} exist, 
is the specific form of the effective potential
in the gauge case: the condensate interaction terms
originate from a single gauge-fermion term (\ref{fkresult}), and their  
coupling constant is determined by $K$. In contrast, the 
adjoint Higgs model is characterised by an independent 
coupling for the Higgs interactions~\cite{adjoint}, and for certain 
regions of the parameters a change in symmetry occurs.

\begin{centering}
\begin{figure}[htb]
\epsfxsize=2in
\vspace{2cm}

\centerline{\rotate{\rotate{\rotate{\epsffile{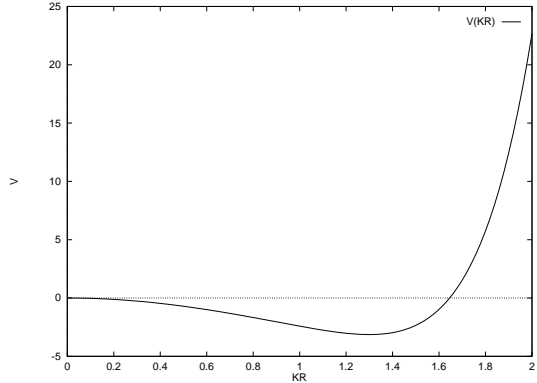}}}}}

\vspace{1cm}
\caption{{\it The effective potential (\ref{smally}) in the case
$\beta_1=0$, $\beta_2=\infty$. The system is in the 
broken-$SU(2)$-symmetry phase for any $K$.}}
\label{fig2}
\end{figure}
\end{centering}

\pr
The incorporation of the gauge interactions 
will change the situation, and induce
non-trivial dynamics which may result in a change in symmetry
for some regions of the gauge coupling constants. 
From the previous result (\ref{smally}), and 
the discussion in section 2, it becomes clear 
that for weak $SU(2)$ and strong enough 
$U_S(1)$
the $SU(2)$ symmetry is broken down to a $U(1)$ subgroup~\cite{farak}. 
The non-trivial issue here 
is whether there exist critical (inverse) couplings $\beta_i^c, i=1,2$, 
above which  the symmetry is restored. 
Let us first concentrate on the axis 
$\beta_2=\infty$, $\beta_1={\rm free}$.  
According to earlier analyses,
either in the continuum or on the lattice~\cite{app,maris,kocic},
there appears to be 
a {\it critical coupling} $\beta_1^c$ 
on this axis above which dynamical mass generation
due to the $U_S(1)$ group cannot take place. This is depicted in 
figure 1.
\pr
The situation concering the $\beta_2$ coupling is more complicated. 
Let us first 
concentrate on the region of strong 
$SU(2)$, $\beta_2 = 0$, keeping $U_S(1)$ 
arbitrary (bottom horizontal axis of fig. 1). 
In this part of the phase diagram one 
can {\it integrate out} the (strongly coupled) $SU(2)$ gauge fields
to derive an effective action for the 
fermion and $U_S(1)$ gauge fields.
The $SU(2)$ path integration 
is performed along the lines of ref. \cite{kawamoto}.
In the strong coupling limit for $SU(2)$, $\beta_2 =0$,
the effective action,
obtained after integration 
of the $SU(2)$ gauge fields, reduces to the a sum of {\it 
one-link contributions}, $S_{eff}=S_{eff}(A,{\overline A})$,
with 
\be 
    A_\mu(x)^a_b = {\overline \Psi}_b(x+a) \gamma_\mu 
U^\dagger_{x,\mu}\Psi ^a (x) 
\qquad  {\overline A}_\mu(x)^a_b = 
{\overline \Psi}_b(x) (-\gamma_\mu) U_{x,\mu}\Psi ^a (x+ a) 
\label{AAbar}
\ee
where $U_{x,\mu}$ denotes the $U_S(1)$ group element, 
$a$ is the lattice spacing, and the latin indices $a,b$  
are colour $SU(2)$ indices. For the $SU(N)$ case 
the quantity $exp(-S_{eff})$ is known
in an expansion over 
$A$, ${\overline A}$~\cite{kawamoto}. This will be sufficient for our 
purposes here: 
\be
S_{eff}= \frac{1}{N}Tr({\overline A}A) 
+ \frac{1}{2N(N^2-1)}
[-Tr[({\overline A}A)^2] + \frac{1}{N}(Tr[{\overline A}A])^2]
+ \dots + \frac{1}{N!}(detA + det{\overline A}) + \dots 
\label{effAA}
\ee
The determinant terms are associated with baryonic states~\cite{kawamoto}.
We also note that 
for the $U(N)$ case the determinant terms are {\it absent}. 
In the phase diagram of fig. 1 the $U(2)$ case 
occurs at the point $\beta_2 \rightarrow 0, \beta_1 \rightarrow 0$. 
In our discussion below we shall approach that point 
asymptotically, by working on the $\beta_2=0$ line, and 
assuming $\beta_1 \ne 0$. 
We first notice that the Abelian phase factors of the $U_S(1)$ interactions
{\it cancel} from the expressions for the traces 
of $A$, ${\overline A}$ in the effective action (\ref{effAA}).
Moreover, from the discussion 
of section 2, we know that 
the $SU(2)$ (strong-coupling) integration  
{\it cannot produce} a parity-invariant 
condensate, since the latter is not an $SU(2)$ singlet~\cite{farak}. 
The resulting effective action should be expressible in terms 
of $SU(2)$ invariant fields. 
Thus, on the axis $\beta_2=0$ there is {\it no 
possibility} for the $U_S(1)$ group to generate a fermion condensate. 
This implies that for very strong $SU(2)$ group the
symmetry is {\it restored} for arbitrary $U_S(1)$ couplings. 
This observation, together with the fact that for weak $SU(2)$ 
couplings there is dynamical formation of parity invariant
fermion condensates along the axis $\beta_1=0$~\cite{app,dor,farak,fm},
implies the existence of a {\it critical } $\beta_2^c$ above which 
the $SU(2)$ symmetry is restored. The issue is whether this critical 
coupling is {\it finite} or it occurs at {\it infinity}.
This issue requires proper lattice simulations, which fall beyond the 
scope of the present work. However in certain specific models, 
like the one in ref. \cite{fm}, this issue might be resolved, as we shall 
discuss in section 5. 
\pr
The conclusion of the above analysis, therefore, is that 
there exists a {\it critical line} in 
the phase diagram of fig. 1, which separates the $SU(2)$-broken 
phase of the theory from the phase where the symmetry is restored.
Its precise shape is conjectural at this stage and can only be determined
by finite $\beta_2$ calculations. This is in progress.
\pr
Having discussed the situation concerning the 
parity-invariant condensate, one should now examine the possibility
of the parity-violating one ($SU(2)$ singlet, ${\cal A}_4 \equiv
\phi$) which might be
generated by the non-Abelian group. 
However, for the case of a {\it single fermion flavour}, $N_f=1$, 
which is the case we have been concentating so far, 
this is not possible. {\it Energetics} 
prohibits the dynamical generation of the $SU(2)$-singlet 
parity-violating condensate~\cite{Vafa,app},
except in the case where there exist 
non-zero parity-violating 
source terms in the lattice action~\cite{ambjorn}. 
This is consistent with the form
of the effective action (\ref{effAA}). Indeed, 
the trace terms in (\ref{effAA}) depend only on the combination 
$A {\overline A}$ which is parity invariant. 
Moreover, 
as can be easily seen, 
the determinant terms do not contain $\phi$:
\bea 
&~& detA + det{\overline A}= -\frac{K^2}{4}[
{\overline \Psi}_{1,\alpha}(x)
{\overline \Psi}_{2,\delta}(x)(\gamma_{\mu})^\alpha_\beta (\gamma ^{\mu})^\delta_\eta \Psi_1^\beta (x + {\hat \mu})\Psi_2^\eta (x+{\hat \mu})
e^{2i\theta_{x,\mu}} - \nn \\
&~& {\overline \Psi}_{2,\delta}(x)
{\overline \Psi}_{1,\alpha} (x)(\gamma_{\mu})^\delta_\beta 
(\gamma ^{\mu})^\alpha_\eta \Psi_1^\beta (x+{\hat \mu})
\Psi_2^\eta (x+{\hat \mu})e^{2i\theta_{x,\mu}}
+ \nn \\
&~& {\overline \Psi}_{1,\alpha}(x+{\hat \mu})
{\overline \Psi}_{2,\delta} (x+{\hat \mu})(\gamma_{\mu})^\alpha_\beta (\gamma ^{\mu})^\delta_\eta \Psi_1^\beta (x)\Psi_2^\eta (x)e^{-2i\theta_{x,\mu}}
- \nn \\
&~& {\overline \Psi}_{2,\delta} (x+{\hat \mu})
{\overline \Psi}_{1,\alpha} (x+{\hat \mu})(\gamma_{\mu})^\delta_\beta 
(\gamma ^{\mu})^\alpha _\eta \Psi_1^\beta (x)\Psi_2^\eta 
(x)e^{-2i\theta_{x,\mu}}]
\label{detAterms}
\eea
where ${\hat \mu}$ denotes the unit step in the $\mu$ direction, 
Greek indices are spinor indices, $1,2$ denote colour $(SU(2))$ indices,
and $\theta_{i,\mu}$ is the $U_S(1)$ gauge phase.
This expression depends on the baryonic states~\cite{kawamoto} 
${\cal B}_{\alpha\beta} \propto \epsilon_{ab}\Psi_{\alpha}^a\Psi _{\beta}^b$,
$\alpha,\beta$ spinor indices, $a,b$ colour indices, and, hence,  
does not contain the parity-violating ${\cal A}_4$ meson state
(\ref{singlets}) ${\cal A}_4 \sim {\overline \Psi}_1\Psi_1
+ {\overline \Psi}_2\Psi_2$. Thus, the form of the effective action 
(\ref{effAA}) in our case
is consistent with the impossibility of the 
spontaneous breaking of parity in vector-like theories in  
odd dimensions~\cite{Vafa}.
\pr
A final point concerns the {\it multiflavour} 
fermionic case, $N_f > 1$.
In the case of 
an {\it even} flavour 
number it is {\it possible} for a strong $SU(2)$ group to generate 
a {\it parity invariant combination} of fermion masses, if 
$N_f/2$ flavours get positive masses and $N_f/2$ get
masses equal in magnitude but opposite in sign~\cite{app,Vafa}.
This completes our (preliminary) study of the 
phase diagram (fig. 1) of the $SU(2) \times U_S(1)$
model of a gauged chiral symmetry considered above. 

\section{Superconducting Properties}

As a final topic of our generic analysis 
of three-dimensional gauge models we would like to discuss 
the superconducting consequences of 
the above dynamical breaking patterns of the $SU(2)$ group. 
{\it Superconductivity} is obtained upon coupling 
the system to external elelctromagnetic potentials, 
which leads to the presence of an additional gauge-symmetry, 
$U_E(1)$, that of ordinary electromagnetism.

\begin{centering}
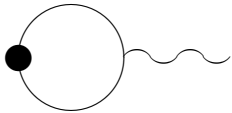
\begin{figure}[htb]
\vspace{2cm}
%
%\begin{figure}[h]
\bigphotons
%\label{vacpol3}
\begin{picture}(30000,5000)(0,0)
\put(20000,0){\circle{100000}}
%\drawline\gluon[\E\REG](22000,0)[2]
\drawline\photon[\E\REG](22000,0)[4]
\put(18000,0){\circle*{1000}}
\end{picture}
%\end{figure}
%\centerline{\epsffile{fig2.eps}}
%
\vspace{1cm}
\caption{{\it Anomalous one-loop Feynman matrix element,
leading to a Kosterlitz-Thouless-like breaking of the 
electromagnetic $U_{E}(1)$ symmetry, and thus 
superconductivity, once a fermion 
mass gap opens up. The wavy line represents the $SU(2)$ 
gauge boson $B_\mu^3$,
which remains massless, while the blob denotes an insertion 
of the fermion-number
current  $J_\mu={\overline \Psi}\gamma_\mu \Psi$.
Continuous lines represent fermions.}}
\label{fig3}
\end{figure}
\end{centering}

\pr
Upon the 
opening of a mass gap in the fermion (hole) spectrum, one obtains 
a non-trivial result for the 
following
Feynman matrix element: 
${\cal S}^a = <B^a_\mu|J_\nu|0>,~a=1,2,3$, with $J_\mu ={\overline
\Psi}\gamma _\mu \Psi $, the fermion-number current.
Due  
to the colour-group structure, only the massless $B^3_\mu $ 
gauge boson of the $SU(2)$ group, corresponding to the $\sigma _3$
generator in two-component notation, contributes to the 
matrix element. 
The non-trivial result for the matrix element ${\cal S}^3$
arises from an {\it anomalous one-loop graph}, depicted in 
figure 3, and it
is given by~\cite{RK,dor}:
\be
    {\cal S}^3 = <B^3_\mu|J_\nu|0>=({\rm sgn}{M})\epsilon_{\mu\nu\rho}
\frac{p_\rho}{\sqrt{p_0}} 
\label{matrix2}
\ee
where $M$ is the parity-conserving fermion mass 
(holon condensate), generated dynamically 
by the $U_S(1)$ group. As with the 
other Adler-Bell-Jackiw anomalous graphs in gauge theories, 
the one-loop result (\ref{matrix2}) 
is {\it exact} and receives no contributions from higher loops~\cite{RK}. 
\pr
This unconventional 
symmetry breaking (\ref{matrix2}), 
does {\it not have a local order parameter}~\cite{RK,dor},
since the latter is inflicted by strong phase fluctuations, 
thereby resembling  the
Kosterlitz-Thouless
mode of symmetry breaking~\cite{KT}. The {\it massless} Gauge Boson 
$B_\mu^3$ of the 
unbroken $\sigma_3-U(1)$ subgroup of $SU(2)$ is responsible for the 
appearance of a {\it massless pole} in the electric current-current 
correlator~\cite{dor}, which is the characteristic feature 
of any {\it superconducting theory}. As discussed in ref. \cite{dor},
all the standard properties of a superconductor, such as 
the Meissner
effect, infinite conductivity, flux quantization, London action etc. are 
recovered in such a case. 
The field $B^3_\mu$, or rather its {\it dual} $\phi$ defined by
$\partial _\mu \phi \equiv \epsilon_{\mu\nu\rho}\partial_\nu B^3_\rho$,
can be identified with the Goldstone
Boson of the broken $U_{em}(1)$ (electromagnetic) symmetry~\cite{dor}.

\section{Application to Doped Planar Antiferromagnets}

Before closing we would like 
to discuss how these results
can be connected with the low-energy limit 
of {\it doped antiferromagnetic} planar systems
of
relevance to the physics of high-temperature
superconductors. We shall be very brief in our discussion here.
For more details we refer the reader to ref. \cite{fm}
and references therein. 
The model considered in \cite{fm} was the strong-U Hubbard model,
describing doped antiferromegnets with the constraint 
of no more than one elelctron per lattice site.
The key suggestion in ref. \cite{fm}, which lead to 
the non-abelian gauge symmetry structure for the doped antiferromagnet,
was the {\it slave-fermion} spin-charge separation ansatz 
for physical electron operators 
at {\it each lattice site} $i$~\cite{fm}:
\be
\chi _{\alpha\beta,i} 
\equiv \left(
\begin{array}{cc}
c_1 \qquad c_2 \\
c_2^\dagger \qquad -c_1^\dagger \end{array}
\right)_i \equiv {\widehat \psi} _{\alpha\gamma,i}{\widehat z}_{\gamma\beta,i} =
\left(\begin{array}{cc}
\psi_1 \qquad \psi_2 \\
-\psi_2^\dagger \qquad \psi_1^\dagger \end{array}
\right)_i~\left(\begin{array}{cc} z_1 \qquad -{\overline z}_2 \\
z_2 \qquad {\overline z}_1 \end{array} \right)_i 
\label{ansatz2}
\ee
where 
$c_\alpha$, $\alpha=1,2$ are electron anihilation
operators, the Grassmann variables $\psi_i$, $i=1,2$ 
play the r\^ole of holon excitations, while the bosonic
fields $z_i, i=1,2,$ represent magnon excitations~\cite{Anderson}.
The ansatz (\ref{ansatz2}) 
has spin-electric-charge separation, since only the 
fields $\psi_i$ carry {\it electric} charge.
This ansatz characterizes the proposal 
of ref. \cite{fm} for the dynamics underlying 
{\it doped} antiferromagnets.
In this context, the 
holon fields ${\widehat \psi} _{\alpha\beta}$ 
may be  
viewed as substructures of the 
physical electron $\chi_{\alpha\beta}$~\cite{quark},
in close analogy to the `quarks' of $QCD$.   
\pr
As argued in ref. \cite{fm} 
the ansatz is characterised by   
the following {\it local}
phase (gauge) symmetry structure: 
\be
  G=SU(2)\times U_S(1) \times U_E(1) 
\label{group}
\ee
The 
local 
SU(2) symmetry is discovered if one defines the transformation 
properties of the ${\widehat z}_{\alpha\beta}$ 
and ${\widehat \psi} ^\dagger_{\alpha\beta}$ 
fields to be given by left multiplication
with the $SU(2)$ matrices, and pertains to the spin degrees of freedom.
The 
local $U_S (1)$ `statistical' phase symmetry, which 
allows fractional statistics of the spin and charge 
excitations. This is an exclusive feature
of the three dimensional geometry, and is similar in spirit
to the bosonization technique of the spin-charge 
separation ansatz of ref. \cite{marchetti},
and allows the alternative possibility 
of representing the holes as slave bosons and  
the spin excitations as fermions. 
Finally the $U_E(1)$ symmetry is due to the electric
charge of the holons. 
\pr
The pertinent long-wavelength gauge model, describing the low-energy 
dynamics
of the large-U Hubbard 
antiferromagnet, in the spin-charge separation phase (\ref{ansatz2}),
assumes the form~\cite{fm}: 
\bea
&~&H_{HF}=\sum_{<ij>} tr\left[(8/J)\Delta^\dagger_{ij}\Delta_{ji}
+ K(-t_{ij}(1 + \sigma_3)+\Delta_{ij}){\widehat \psi}_j V_{ji}U_{ji}
{\widehat \psi}_i^\dagger\right] + \nn \\ 
&~&\sum_{<ij>}tr\left[ K{\overline {\widehat z}}_iV_{ij}U_{ij}{\widehat z}_j\right] + h.c. 
\label{Hub}
\eea
where $J$ is the Heisneberg antiferromagnetic interaction, 
$K$ is a normalization constant, and 
$\Delta_{ij}$ is a Hubbard-Stratonovich field that linearizes
four-electron interaction terms in the original Hubbard model, 
and 
$U_{ij}$,$V_{ij}$ are the link variables for the $U_S(1)$ and 
$SU(2)$ groups respectively. 
The conventional lattice gauge theory form of the action (\ref{Hub}) 
is derived upon freezing the fluctuations of the $\Delta_{ij}$ 
field~\cite{fm}, and   
integrating out the magnon fields, $z$,
in the path integral. This latter operation yields 
appropriate Maxwell kinetic terms for the link variables 
$V_{ij}$, $U_{ij}$, 
in a low-energy derivative expansion
~\cite{IanA,Polyakov}.
On 
the lattice such kinetic terms 
are given by plaquette terms of the form~\cite{fm}:
\be   
\sum_{p} \left[\beta_{SU(2)}(1-Tr V_p) + \beta_{U_S(1)}(1-Tr U_p)\right]
\label{plaquette} 
\ee
where $p$ denotes sum over 
plaquettes of the lattice, 
and $\beta_{U_S(1)} \equiv \beta_1 $, $\beta_{SU(2)} \equiv \beta_2 = 
4\beta_1$ 
are the dimensionless (in units of the lattice spacing) 
inverse square couplings of the $U_S(1)$ and $SU(2)$ groups,  
respectively~\cite{fm}. The above relation between the $\beta_i$'s
is due to the specific form of the $z$-dependent terms in (\ref{Hub}),
which results in the same induced couplings $g_{SU(2)}^2=g_{U_S(1)}^2$.
Moreover,  there is a non-trivial connection of the 
gauge group couplings to $K$~\cite{fm}: 
\be
K \propto g_{SU(2)}^2 = g_{U_S(1)}^2 \propto J\eta
\label{connection}
\ee 
with $\eta$ the doping concentration in the sample~\cite{fm,dorstat}. 
To cast the 
symmetry structure in a 
form that is familiar to 
particle physicists, one may change representation 
of the $SU(2)$ group, and instead of working with $2 \times 2$ 
matrices in (\ref{ansatz2}), one may use a representation 
in which the fermionic matrices ${\widehat \psi}_{\alpha\beta}$ 
are represented as two-component (Dirac) spinors in `colour' space:
\be
{\tilde \Psi}_{1,i}^\dagger =\left(\psi_1~~-\psi_2^\dagger\right)_i,~~~~
{\tilde \Psi} _{2,i}^\dagger=\left(\psi_2~~\psi_1^\dagger
\right)_i,~~~~~i={\rm Lattice~site} 
\label{twospinors}
\ee
In this representation  
the two-component spinors ${\tilde \Psi} $ (\ref{twospinors}) 
will act as Dirac spinors, and the $\gamma$-matrix (space-time)
structure will be spanned by the irreducible 
$2 \times 2$ representation. 
By 
assuming a background $U_S(1)$ 
field of flux $\pi$ per lattice plaquette~\cite{dor},
and considering quantum fluctuations around this background
for the $U_S(1)$ gauge field, 
one can show that there is a Dirac-like structure 
in the fermion spectrum~\cite{Burk,AM,dor,dorstat}, 
which leads to a conventional 
Lattice gauge theory form for the effective low-enenrgy Hamilonian of the
large-$U$,  doped Hubbard model~\cite{fm}. Remarkably, 
this lattice gauge theory has the 
same form as (\ref{effeaction}). The constant $K$ of (\ref{effeaction}) can then be identified
with $K$ in (\ref{Hub}).  

In the above context, a strongly coupled
$U_S(1)$ group can dynamically generate a mass gap 
in the holon spectrum, which breaks the $SU(2)$ local symmetry
down to its Abelian subgroup generated by the $\sigma_3$ matrix.  
{}From the view point of the statistical model (\ref{Hub}), 
the breaking of the $SU(2)$ symmetry down to its Abelian 
$\sigma_3$ subgroup may be interpreted as  
restricting the holon hopping effectively to 
a single sublattice, since 
the intrasublattice hopping is suppressed 
by the mass of the gauge bosons.  
In a low-energy
effective theory of the massless degrees of freedom 
this reproduces the 
results of ref. \cite{dor,Sha}, derived 
under a large-spin 
approximation for the antiferromagnet, $S \rightarrow \infty$, 
which is not necessary 
in the present approach. 
\pr
We now remark 
that, since 
$K$ 
is proportional
to
the doping concentration in the sample~\cite{dorstat,fm}, 
$K \propto J\eta$, then, 
the phase diagram of fig. 1 
indicates
the existence, in general,  
of an {\it upper} and a {\it lower} 
bound for $\eta$ in order to 
have superconductivity in the 
model~\footnote{The relation (\ref{connection}) 
also implies that the constant $K$ 
has a non-trivial physical significance,
and one should not absorb it in a redefinition 
of the fermion fields. In terms of the statistical model,
there are higher interactions among the fermions 
which render such a redefinition 
not appropriate.}. 
These bounds 
correspond to
the points A and B, respectively, 
at which the straight line $\beta_2=4\beta_1$
intersects the critical line of the phase diagram of fig. 1. 
The existence of the lower bound in the doping concentration 
would imply that 
in 
planar antiferromagnetic models
antiferromagnetic order is destroyed,
in favour of superconductivity, 
{\it above} a critical doping concentration. 
This point of view seems to be supported
by preliminary results of 
lattice simulations~\cite{hirsch}. 
The upper bound, if exists, 
would also be of interest, since it is known 
that in the high-$T_c$ 
cuprates superconductivity is destroyed
above a doping concentration of a few per cent. 
However, 
in view of the result (\ref{smally}), the
relation (\ref{connection}), that characterizes the 
coupling constants of the effective model (\ref{Hub}), 
probably implies that the 
inverse critical 
coupling, $\beta^c_2$,  
below which
the $SU(2)$ symmetry is restored, 
occurs at zero: $\beta_2^c\rightarrow 0$
(dashed line in fig. 1). In that case there will be no finite 
upper bound for the 
doping concentration coming from the phase diagram of fig. 1. 
More detailed 
investigations along this line of thought, 
and a quantitative study of the effects of doping 
in the context of a renormalization group analysis  
for the model (\ref{Hub}) are in progress.

\paragraph{}
\noindent {\Large {\bf Acknowledgements}}
\paragraph{}
The authors would like to thank 
I. Aitchison and G. Koutsoumbas
for 
discussions. They also wish to thank the CERN Theory 
division for hospitality during the last stages of this work. 
K.F. wishes to acknowledge partial financial support 
from PENED 95 Program, No. 1170, of the Greek General Secretariat
of Research and Technology.

\end{document}